\documentclass[final,1p,times]{elsarticle} 
\biboptions{sort&compress}
\usepackage{hyperref}

\usepackage[outdir=./]{epstopdf}

\usepackage{amssymb}

\usepackage{lineno}
\usepackage{siunitx}

\journal{a peer-reviewed journal}

\begin{document}

\begin{frontmatter}


\title{Drift in a Popular Metal Oxide Sensor Dataset Reveals Limitations for Gas Classification Benchmarks}

\author[inst1,inst2]{Nik Dennler}
\author[inst2,inst1]{Shavika Rastogi}
\author[inst3,inst4]{Jordi Fonollosa}
\author[inst2]{André van Schaik}
\author[inst1]{Michael Schmuker}

\affiliation[inst1]{organization={UH Biocomputation Research Group, Centre for Computer Science and Informatics Research, University of Hertfordshire},
            city={Hatfield},
            country={United Kingdom}}

\affiliation[inst2]{organization={International Centre for Neuromorphic Systems, Western Sydney University},
            city={Sydney},
            country={Australia}}

\affiliation[inst3]{organization={Departament d’Enginyeria de Sistemes, Automàtica i Informàtica Industrial, Universitat Politècnica de Catalunya},
            city={Barcelona},
            country={Spain}}
            
\affiliation[inst4]{organization={Institut de Recerca Sant Joan de Déu},
            city={Esplugues de Llobregat},
            country={Spain}}

\begin{abstract}
Metal Oxide (MOx) electro-chemical gas sensors are a sensible choice for many applications, due to their tunable sensitivity, their space-efficiency and their low price. 
Publicly available sensor datasets streamline the development and evaluation of novel algorithm and circuit designs, making them particularly valuable for the Artificial Olfaction / Mobile Robot Olfaction community. In 2013, Vergara et al.\ published a dataset comprising 16 months of recordings from a large MOx gas sensor array in a wind tunnel, which has since become a standard benchmark in the field.
Here we report a previously undetected property of the dataset that limits its suitability for gas classification studies.
The analysis of individual measurement timestamps reveals that gases were recorded in temporally clustered batches. The consequential correlation between the sensor response before gas exposure and the time of recording is often sufficient to predict the gas used in a given trial. Even if compensated by zero-offset-subtraction, residual short-term drift contains enough information for gas classification. We have identified a minimally drift-affected subset of the data, which is suitable for gas classification benchmarking after zero-offset-subtraction, although gas classification performance was substantially lower than for the full dataset.
We conclude that previous studies conducted with this dataset very likely overestimate the accuracy of gas classification results. 
For the 17 potentially affected publications, we urge the authors to re-evaluate the results in light of our findings. Our observations emphasize the need to thoroughly document gas sensing datasets, and proper validation before using them for the development of algorithms. 

\end{abstract}




\end{frontmatter}


\section{Introduction}

\label{sec:introduction}
Over the last 50 years, artificial olfaction has evolved from an almost niche field of study into a thriving interdisciplinary research area. Many use cases have been addressed, for example the detection of hazardous gases or pollutants \cite{stetter1986detection}, spoilage localization \cite{maier2006monitoring}, mobile olfactory robotics \cite{s6111616}, health monitoring \cite{8846091} and medical screening \cite{loizeau2013piezoresistive}; and artificial olfaction is expected to address many more use cases in the future \cite{Covington.2021}. 
A key challenge in artificial olfaction is to identify a range of  odorants at high specificity. One way to achieve this is to use an array of multiple gas sensors, each with a rather large selectivity and low specificity, and extract the identity of the presented odor using pattern recognition. Metal Oxide (MOx) electro-chemical gas sensors are a widely used candidate for such sensor arrays. Their sensing layer can be tuned to different analyte classes and they are very cost- and space efficient since they require little electronic periphery. One big drawback of MOx sensors is their susceptibility to sensor drift---the gradual and unpredictable variation of signal response over time when exposed to identical analytes under the same conditions \cite{ZIYATDINOV2010460}. Drift is mostly due to chemical and physical interactions on the sensor site, such as sensor aging (reorganization of the sensor surface over time) and sensor poisoning (irreversible or slowly reversible binding of previously measured gases or other contamination).  Environmental effects such as changes in humidity, temperature or pressure also affect the sensor response. In order to successfully overcome drift, it is essential to carefully craft the experimental procedure accordingly, for example by randomizing the order of analytes presented to avoid any correlation between gas identity and sensor drift.

Setting up an electronic olfaction system still requires custom design of electronics and data analysis systems. Therefore, many users, developers and researchers of the Artificial Olfaction / Mobile Robot Olfaction (AO/MRO) community will initially look for previously recorded datasets. A number of datasets are publicly available, covering a range of tasks and use cases \cite{s100100036, VERGARA2013462, VERGARA2012320, s141019336, FONOLLOSA2015618, ZIYATDINOV2015538,  FONOLLOSA20161044, BURGUES201813, RODRIGUEZGAMBOA2019104202}. One of the most popular datasets contains MOx sensor data sampled in a wind tunnel, for different gases and different experimental parameters, over a time of 16 months \cite{VERGARA2013462, VergaraDataset}. This dataset by Vergara et al.\ has been cited more than 100 times\footnote{According to Google Scholar as of July 2021}. In at least 17 publications it has been used as a benchmark for gas classification algorithms \cite{Battaglino_2018, Vervliet_2016, Imam_2020, Choi_2019, Zhou_2019, Monroy_2016, Fan_2018, Rodriguez_Gamboa_2021, Zhou_2019_, Kolda_2020, Mishra_2018, Coudron2019, Araujo_2019, VervlietPHD, MonroyPHD, gugel2016machine, chang2020uci}. It has also been used for gas source location estimation \cite{Schmuker_2016, BurguesPHD, Burgues_2019, Burgu_s_2020} and other applications \cite{Lee_2017, Monroy_2017, Schneider_2018, Mitchell_2020, Cardellicchio_2019, gilman2021grassmannian}.

Here, we reveal a fundamental limitation of the dataset introduced above. First, we observed that gases were not presented in random order, but in distinct batches, sometimes recorded weeks or months apart. In consequence, the sensor recordings are contaminated by slow baseline drift effects that correlate with time, and therefore with gas identity. We show that since both the gas identity and sensor baseline correlate with time, it is possible to identify trials using a specific gas only by looking at the baseline response, before any gas has been released. In addition, we show that, even after correcting for slow drift by subtracting the average of the first few sensor readings of each experimental trial, residual short-term drift effects are characteristic enough to identify trials where specific gases have been used, using the baseline alone. Moreover, when further minimizing the impact of drift by selecting the least-affected subset of recordings and compensating for drift as much as possible, the gas classification performance is far inferior to the numbers we obtained when using the full dataset. Therefore we conclude that this dataset is only of limited use for gas classification benchmarking, and that previously reported classification results based on this dataset are likely severely overoptimistic. Finally, we give a perspective on how the measurement protocol could be improved to mitigate this problem, and elaborate on what tasks the dataset can be appropriately used for, i.e., tasks that are not affected by the drift contamination.

\section{Dataset}
\begin{table}[t]
\centering
\begin{tabular}{|c|c|c|c|c|}
\cline{1-2} \cline{4-5}
\textbf{Col. no.} & \textbf{Sensor model} &  & \textbf{Col. no.} & \textbf{Sensor model} \\ \cline{1-2} \cline{4-5} 
1                 & TGS 2611              &  & 5                 & TGS 2600              \\ \cline{1-2} \cline{4-5} 
2                 & TGS 2612              &  & 6                 & TGS 2600              \\ \cline{1-2} \cline{4-5} 
3                 & TGS 2610              &  & 7                 & TGS 2620              \\ \cline{1-2} \cline{4-5} 
4                 & TGS 2602              &  & 8                 & TGS 2620              \\ \cline{1-2} \cline{4-5} 
\end{tabular}
\caption{Metal Oxide (MOx) sensors included in each 8-sensor array. All sensors were manufactured by Figaro USA, Inc\cite{figaro}.}
\label{tab:sensors}
\end{table}
The dataset in question \cite{VERGARA2013462} consists of 18000 times-series measurements recorded from a 72 MOx gas sensor array-based chemical detection platform, in which 10 different analyte gases (Acetone, Acetaldehyde, Ammonia, Butanol, Ethylene, Methane, Methanol, Carbon monoxide, Benzene, and Toluene) were measured over a period of 16 months. The sensor platform consisting of nine modules, each equipped with eight MOx sensors (see Table \ref{tab:sensors} for sensor types), was placed in a $\SI{2.5}{\meter} \times \SI{1.2}{\meter} \times \SI{0.4}{\meter}$ wind tunnel, at six different distances, normal to the wind direction (see Figure \ref{fig:protocol}a for a schematic). Each sensor module was integrated with a sensor controller, which enabled data collection at 12-bit resolution and a sampling rate of $\SI{100}{\hertz}$. Computer-supervised mass flow controllers, a multiple-step motor-driven exhaust fan and a flat, non-inclined floor ensured a sheer, yet turbulent chemical air stream across the wind tunnel. 
Different experimental conditions were tested, namely three different wind speeds set by the fan ($\SI{0.1}{\meter \per \second}$, $\SI{0.21}{\meter \per \second}$, $\SI{0.34}{\meter \per \second}$) and five different sensor operating voltages ($\SI{4.0}{\volt}$, $\SI{4.5}{\volt}$, $\SI{5.0}{\volt}$, $\SI{5.5}{\volt}$, $\SI{6.0}{\volt}$). Before each measurement, a combination of the experimental parameters \textit{gas}, \textit{location}, \textit{wind speed}, \textit{operating voltage} was  selected, until each combination was repeated 20 times. Each measurement lasted for $\SI{260}{\second}$, where gas was released between $t=\SI{20}{\second}$ and $t=\SI{200}{\second}$. Before and after each experiment, the wind tunnel was ventilated at the maximum speed ($\SI{0.34}{\meter \per \second}$) for two minutes to assert the reestablishment of sensor response baseline. For this analysis, we interpolated and re-sampled the data for dealing with missing data points, and further converted the sensor voltage readings $V_{sensor}$ given in the dataset to sensor resistance values $R_{sensor}$, according to Eq. \ref{eq:resistance}. The readings of sensor 1 for all boards were discarded due to excessive sensor noise. Figure \ref{fig:protocol}b shows the the responses of a sensor board to one gas in a typical trial. If not noted otherwise, we considered the wind tunnel location P4 B5 (wind-downstream from the gas source, see Figure \ref{fig:protocol}a for wind tunnel schematics), as here we expect a high gas exposure.
\begin{equation}
    R_{sensor} = \SI{10}{\kilo \ohm} \times \frac{\SI{3.11}{\volt}-V_{sensor}}{V_{sensor}}
    \label{eq:resistance}
\end{equation}
\section{Results}
\label{sec:results}
\subsection{Non-random order of gas measurements}
\begin{figure*}[t]
    \centering
    \includegraphics[width=0.999\linewidth]{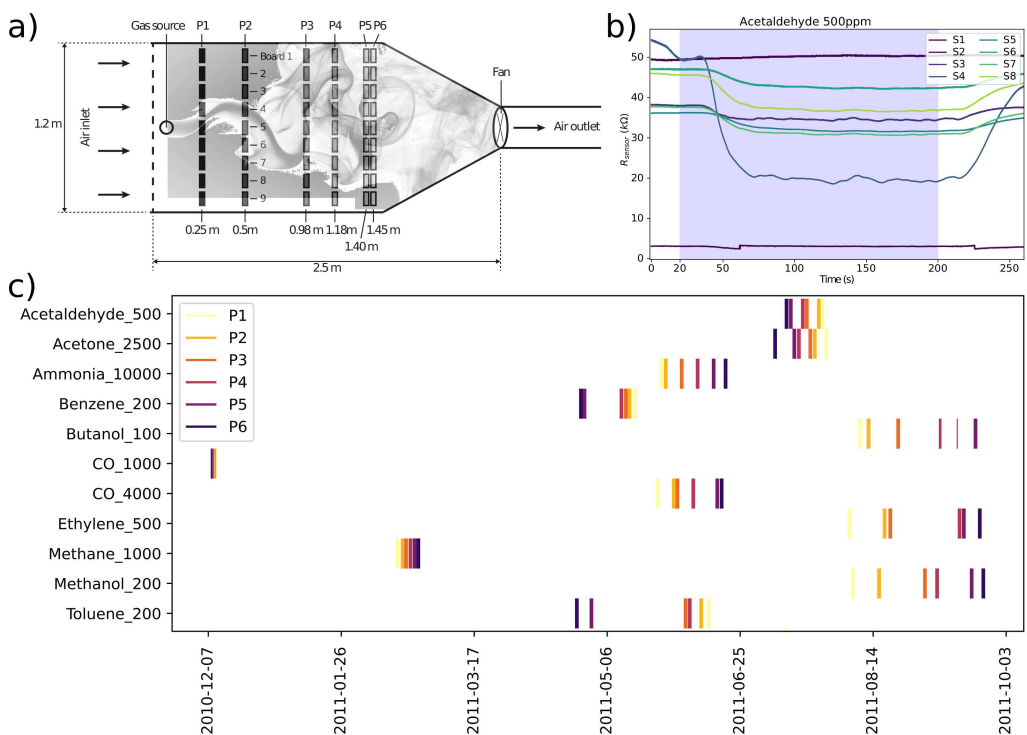}
    \caption{Experimental procedure of Vergara et al. \cite{VERGARA2013462} \textbf{a)} Setup of sensor boards in wind tunnel. Adapted from \cite{VERGARA2013462}. \textbf{b)} Example time series recorded from all sensors on one sensor module (location P4, module 5, Acetaldehyde, \SI{0.21}{\meter\per\second} airflow velocity$, \SI{6}{\volt}$ operating voltage, trial 1). The shaded portion denotes the period during which the analyte was injected into the wind tunnel. \textbf{c)} Event-plot of timestamps for different gas trials. Each vertical line represents 300 trials, which were performed too close to each other for them to be visually distinguishable in this representation. The row name indicates the measured gas and its concentration in parts-per-million ($ppm$). CO at $\SI{1000}{ppm}$ was removed from further analysis since significantly fewer trials were performed than for the other analytes. }
    \label{fig:protocol}
\end{figure*}

The dataset is organized very well; it contains the raw data, which is not common practice, but very useful for checking its validity. The time of recording of individual measurement is encoded as part of the name of the file containing the time series. We extracted the times of recording from the filenames to analyse the temporal order of measurements. Figure \ref{fig:protocol}c shows when measurements have been made, arranged by gas identity and sensor position. It is evident that gases have not been measured in random order, but in distinct batches that cluster in time. Only rarely do measurements of different gases overlap in time; more often, measurement batches are several weeks apart. In no case have gases been alternated on a per-trial basis. In addition, we observe that also other experimental parameters like distance-to-source, wind speed, sensor temperature were selected in an sequential fashion rather than in random order (not shown). 

It should be noted that the iterative, batched arrangement of gas identity and parameter settings is not evident from the description of the dataset provided by the authors, neither in the original paper, nor in the documentation contained in the UCI repository \cite{VergaraDataset}. Describing the experimental protocol, Vergara et al. stated that (quote) \emph{"This measurement procedure was reproduced exactly for each gas category exposure, landmark location in the wind tunnel, operating temperature, and airflow velocity in a random order and up until all pairs were covered."}\cite{VERGARA2013462}. This could be read as to imply that all experimental parameters that define an experiment were selected randomly before each trial, including which gas to release---which would mitigate, to a large extent, the detrimental effect of baseline drift on gas identification benchmarks---which is not the case, as we show here. 

\subsection{Drift in baseline over time}
\begin{figure*}[t]
    \centering
    \includegraphics[width=0.999\linewidth]{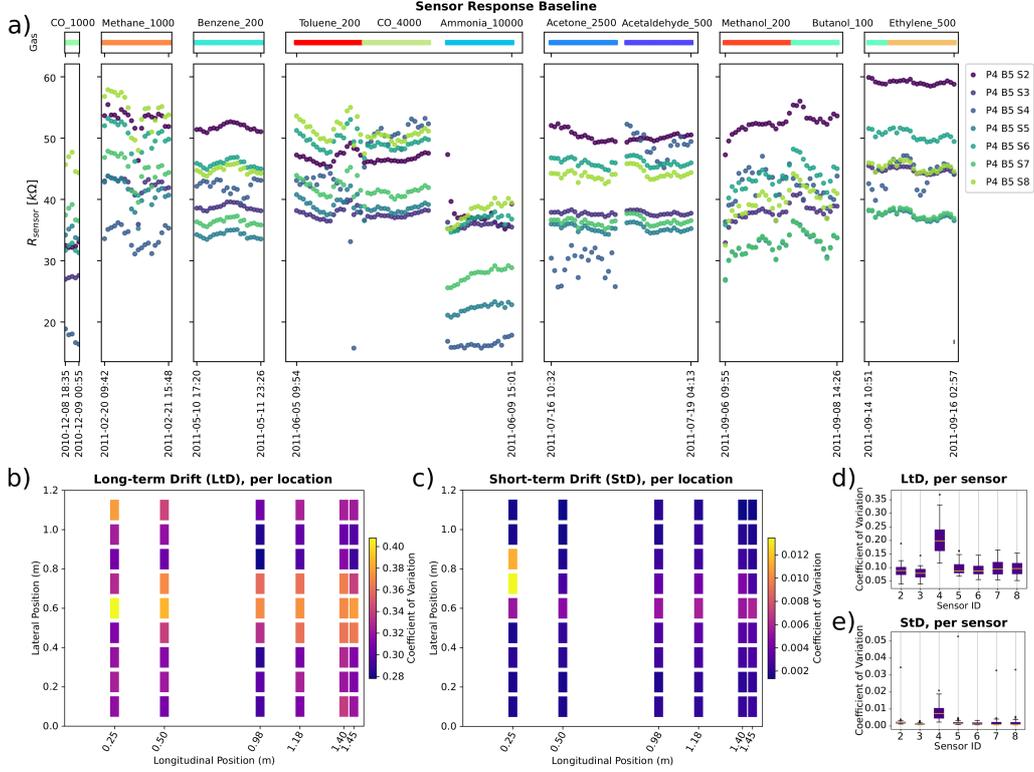}
    \caption{Drift analysis of Vergara et al. dataset \cite{VERGARA2013462}. \textbf{a)} Baseline for each sensor and experimental trial. Dots represent the mean sensor resistance during the time before gas release ($\SI{20}{s}$). Top row indicates the gas and its concentration (in $ppm$) used in the corresponding sessions. \textbf{b)}-\textbf{e)} Spatial drift distribution analysis using the coefficient of variation, for spatial wind tunnel location (\textbf{b)}\&\textbf{c)}) and sensor board (\textbf{d)}\&\textbf{e)}). \textbf{b)}\&\textbf{d)} display the long-term-drift across the whole experimental duration (16 months), where \textbf{c)}\&\textbf{e)} display the within-trial short-term drift. For all experiments shown here, the wind flow speed was fixed at $\SI{0.21}{\meter\per\second}$, while the hotplate voltage was set to $\SI{6}{\volt}$. All ten gases and sensors 2 - 8 are considered. For \textbf{a)}, only location 4 and board 5 are considered (see Figure \ref{fig:protocol}a for wind tunnel schematics).}
    \label{fig:drift}
\end{figure*}
We investigated the sensor baseline across trials, where here we defined baseline as the sensor readings measured before gas is released into the wind tunnel. Figure \ref{fig:drift}a shows the trial-wise average of sensor baseline values at times $t<t_{release}=\SI{20}{\second}$, for a fixed sensor board location, operating temperature and airflow velocity, versus the date of recording. We observed that baseline varies significantly over time. Long-term drift can be observed as significant discontinuities between recording sessions. Since gas presentations were batched, the baseline pattern often correlates with gas identity. In addition, substantial baseline drift can be observed within some recording sessions.  

\subsection{Spatial distribution of baseline drift}
By design, the gas plume does not disperse homogeneously across the wind tunnel, but expresses turbulent flow. Consequently, the gas exposure at the sensor sites varies, which may alter each sensors response differently. Here we investigate how baseline drift is distributed across the wind tunnel and the sensor board. To quantify the drift effects, we calculated the coefficient of variation ($c_v$) of the baseline, for each sensor and for each board location. $c_v$ is given by the fraction between the standard deviation $\sigma$ and the mean $\mu$:
\begin{equation}
    c_v = \frac{\sigma}{\mu}
\end{equation}
We discriminated between long-term drift over the whole duration of the experiment, and short-term drift within single trials. For quantifying long-term drift, $\sigma$ and $\mu$ were computed from the distribution of trial-wise averages of sensor baseline values, thus $c_v$ describes the variation of the baseline across the whole experimental duration. For short-term drift, $c_v$ is the average of the within-trial $\sigma$-to-$\mu$ ratios. We observed a distinct spatial pattern in the distribution of long-term drift coefficients of variation across the wind-tunnel (Figure \ref{fig:drift}b). The long-term drift effects were strongest in sensor boards close to the center line of the wind tunnel, where gas concentration is expected to be highest. This observation may indicate that long-term drift could be caused by exposure to the sample gas. Long-term drift affected all sensors, although sensor 4 was affected most strongly (Figure \ref{fig:drift}c. This sensor is a Figaro \texttt{TGS 2602}, which is targeted towards ``Air pollutants (VOCs, ammonia, H2S)'' according to Figaro's website. 

The coefficients of variation for short-term drift within trials were naturally lower in magnitude but exhibited a similar pattern as observed for long-term drift, both spatially and per-sensor (Figures \ref{fig:drift}d and \ref{fig:drift}e). This indicates that also within-trial drift is highest for those sensors exposed to the highest gas concentrations. 


\subsection{Gas Clustering and Classification}
\begin{figure*}[t]
    \centering
    \includegraphics[width=0.999\linewidth]{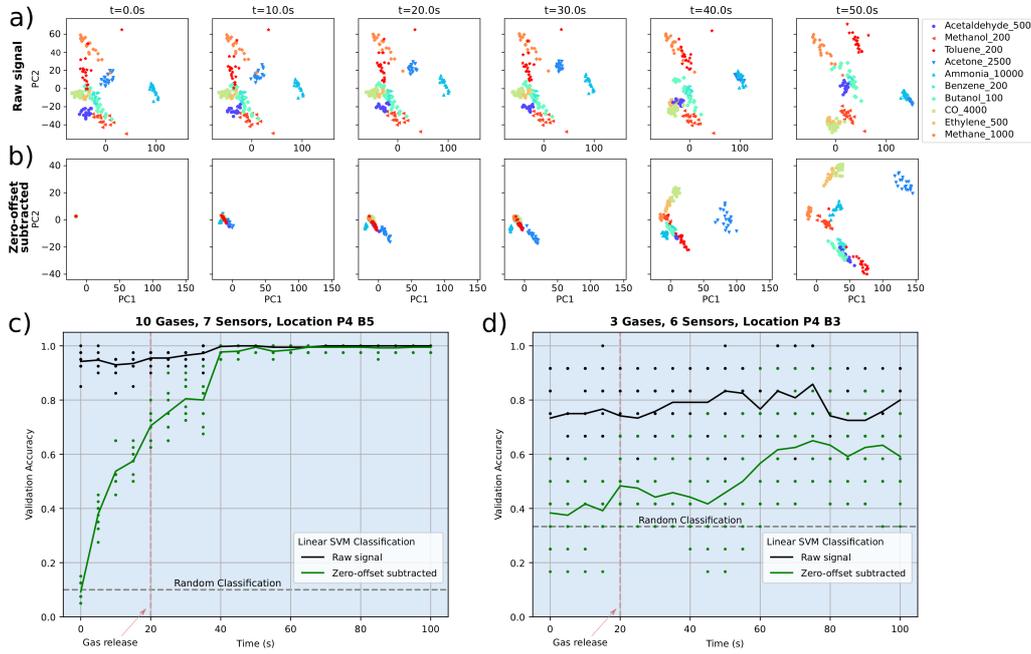}
    \caption{PCA analysis and SVM gas classification of Vergara et al. dataset \cite{VERGARA2013462}. \textbf{a)\&b)} Principal Component Analysis (PCA) of samples within a $\SI{100}{ms}$ time window, at different starting times. Each color and shape corresponds to a different gas. For \textbf{a)}, the raw resistance signal was considered, where for \textbf{b)}, the zero-offset was removed by subtracting the mean resistance in the first $\SI{100}{ms}$ for each sensor. \textbf{c)\&\textbf{d)}} Classification results using a Linear Support Vector Machine classifier. The trials for each gas were randomly split in training and validation datasets with a ratio of $80-20$. Black corresponds to the raw resistance signal, whereas green corresponds to the zero-offset subtracted signal. For all experiments shown here, the wind flow speed was fixed at $\SI{0.21}{\meter\per\second}$, and the hotplate voltage was set to $\SI{6}{\volt}$. For \textbf{a)}-\textbf{c)}, all ten gases and sensors 2 - 8 have been considered, at location 4 and board 5 (see Figure \ref{fig:protocol}a for wind tunnel schematics). For \textbf{d)}, the gases Methanol, Ethylene and Butanol have been considered, measured with sensors 2-3 and 5-8, at location 4 and board 3.}
    \label{fig:classification}
\end{figure*}

Since both the baseline drift and the identity of the gas used in a trial correlate with time, we tested how much information about the gas could be obtained from the baseline signal alone. 

Figure \ref{fig:classification}a shows a Principal Components Analysis (PCA) plot of the raw baseline values for each gas, at a range of times after the start of trials. Each plot presents a snapshot of a $\SI{100}{\milli\second}$ time window. The PCA was computed using all trials in all windows. Each snapshot is a projection of the data into this global space. We observed distinct gas-specific clusters already at $t=\SI{0}{\second}$, before any gas was released into the tunnel at $t_{release}=\SI{20}{\second}$. The clusters change slightly between 30 and 40 seconds, which we assume is when the gas has reached and reacted with the sensor site.

Next, we attempted to compensate for long-term drift effects by subtracting the average of the first $\SI{100}{\milli\second}$ window, i.e.\ for $t\in [\SI{0.0}{\second}, \SI{0.1}{\second})$. The data then only contains the difference of the sensor response relative to the start of a trial. This is a standard procedure when dealing with MOx-sensor data. For Figure \ref{fig:classification}b we computed a global PCA on the data compensated for long-term drift and used the same windows as before to visualize the evolution of sensor responses. By design, at $t=\SI{0}{\second}$ there are no visible clusters. Interestingly, although the zero-baseline has been subtracted, we still observe the formation of clusters before the release of the gas. We interpret this observation as the manifestation of  short-term drift within trials (\emph{cf.} Figures \ref{fig:drift}d and \ref{fig:drift}e). It indicates that short-term drift also changed over time, in a way that correlates with gas identity. 

These observations were confirmed using a time-windowed supervised classification approach with a linear Support Vector Machine (SVM) classifier\footnote{We used a linear kernel with regularization parameter $C=1.0$.}. The classifier was trained and tested separately for each time window. We used a 4-to-1 random training to test split, i.e.\ training on 80\% of the trials in each window and testing using the remaining 20\%, where we repeated the procedure 10 times. Figure \ref{fig:classification}c shows the classifier performance. As expected, the classifier yielded near-perfect gas recognition performance on the raw data (i.e., without compensation for long-term drift), with an average accuracy of $94.3\%$ already on the first time window of a trial, for  $t\in [\SI{0.0}{\second},\SI{0.1}{\second})$. 
Test accuracy increased  slightly for later time windows. From $t=\SI{40}{\second}$ on it converged at $100\%$. We assume that this is when the sensor board is maximally exposed to the gas. 

Strikingly, the classifier also performed well over random choice performance for the baseline-compensated data. While test accuracy was random for the time window at $t=\SI{0}{\second}$, performance was clearly above random already at $t=\SI{5}{\second}$. It increased further to around $80\%$ at $t=\SI{35}{\second}$, before making a step to near $100\%$ at $t=\SI{40}{\second}$. 

Taken together, we observed that the time window before gas exposure contains enough information to identify the gas used in a particular trial, even before the gas has been released into the wind tunnel, and also before the gas likely interacts with the sensor. Baseline compensation for long-term drift mitigates the problem to a certain extent, but there is sufficient information contained in the short-term drift dynamics that allow identification of the gas used in a given trial far above chance level. 

\subsection{Restricted data subset}
Based on our findings in Section \ref{sec:results}.1 - \ref{sec:results}.4, we selected a subset of the data that would be least affected by the drift effects we observed. We selected this subset by three constraints. First, only Methanol, Ethylene and Butanol were considered, since they have been measured within close temporal proximity (see Figures \ref{fig:protocol}c and \ref{fig:drift}a). Second, we removed Sensor 4 from the analysis, as it appears to be particularly affected by drift (see Figures \ref{fig:drift}c \& \ref{fig:drift}e). Third, we used data from sensor board 3 rather than sensor board 5, since our analysis suggested that it was, on average, less affected by drift (see Figures \ref{fig:drift}b \& \ref{fig:drift}d).

We repeated the SVM classification task in this, according to our analysis, less compromised subset. The results are displayed in Figure \ref{fig:classification}d. The classification accuracy for the raw signal is initially still well above chance level at around $75\%$, without changing significantly after gas release. This indicates that long-term drift effects are pronounced enough to enable trial identification even in the restricted dataset. Moreover, classification accuracy increased only very slightly after gas onset. This indicates, paradoxically, that actual gas exposure made little difference for ``gas'' recognition in the restricted dataset.

The picture changed after compensating for long-term drift by subtracting the baseline offset at $t=\SI{0}{\second}$.  Classification accuracy was only slightly above the chance level of $33.3\%$ until gas release.  After gas release, accuracy slowly increased towards to slightly above $60\%$. Therefore, we conclude that the restricted dataset is suitable as a gas classification benchmark when compensating for baseline offset. It should be noted though that a gas recognition accuracy of $60\%$ is much lower than what we and others have reported for the original dataset. On the other hand, the sensor board we selected was located slightly lateral to the downwind axis from the source, therefore likely not as strongly exposed to the gas plume, which potentially affects classification performance negatively (but also apparently reduces sensor drift). 

\section{Discussion}
\label{sec:discussion}
In our analysis we have shown that the different gases have been measured in time-separated batches and not in random order, which makes the data susceptible to sensor drift effects. We have shown that the sensor response baseline correlates with the time of measurement, consistent with long-term drift behaviour. We have also shown that the sensor response baseline alone is enough for `accurate' gas classification, even after compensating for long-term drift by removing the offset at $t=\SI{0}{\second}$. This renders the dataset unsuited for gas classification benchmarks.  

In an attempt to alleviate this limitation, we have identified a subset of the dataset, which, under certain conditions, could be used for gas classification benchmarking. The subset contains three gases that have been measured in close temporal proximity, at a location that appears to be less affected by drift, while disregarding one sensor that is most affected. After applying long-term drift compensation to this subset, we observed what would be expected from a clean experiment: Gas identification is near chance level at the beginning of a trial and rises only after gas has reached the sensor. However, gas classification performance under those conditions is much lower than when using the full dataset, even though the task should be easier due to the smaller number of gases.

We therefore conclude that it is highly likely that many, if not all, of the previous studies using this dataset overestimated the performance of their gas recognition algorithms. This confirms the findings in previous works \cite{Monroy_2016, Rodriguez_Gamboa_2021}, where classification accuracies on this dataset were exceptionally high compared to the other datasets analyzed. We identified at least 17 publications that are potentially affected \cite{Battaglino_2018, Vervliet_2016, Imam_2020, Choi_2019, Zhou_2019, Monroy_2016, Fan_2018, Rodriguez_Gamboa_2021, Zhou_2019_, Kolda_2020, Mishra_2018, Coudron2019, Araujo_2019, VervlietPHD, MonroyPHD, gugel2016machine, chang2020uci}. Since this dataset is one of the most used benchmark dataset for electronic olfaction, it is possible that the state-of-the-art in gas recognition accuracy has been dramatically overestimated. If this is the case (as our study suggests), then it will possibly have hampered progress in the field to a considerable extent, potentially leading to viable approaches to classification being discarded as sub-par performers if tested on other datasets, because they didn't hold up to the overestimated results based on the dataset analyzed here. Thus we believe that, for the field as a whole, our analysis is valuable, since it may enable realistic assessment of gas classification algorithms, and further encourage the collection and sharing of novel gas sensor datasets. 

It should be noted that this dataset, in spite of its limitations, is an excellent example for how datasets should be shared. It contains the raw measurement data and all timestamps of the recordings. This is unfortunately not common practice in the field---often, only derived features are shared. We expressly acknowledge the effort Vergara et al.\ have made to share the data as accurately as possible. Only through their diligence and attention to detail was it possible to identify the underlying problems. 

The dataset still has unique features which make it a tremendous resource for machine olfaction research. It is one of the very few available datasets which have been recorded with a very high temporal resolution in a wind tunnel. Therefore, it includes temporal dynamics of odor concentration which are due to turbulent dispersal. This feature of the dataset has given rise to a study demonstrating that information about source proximity can reliably be extracted from turbulent plumes using metal oxide sensors \cite{Schmuker_2016}, which has been replicated independently \cite{Burgu_s_2020} and confirmed using newly recorded data \cite{Burgues_2019}. Such studies are not affected by the adverse effects discussed in the present study, since they do not attempt to identify odorants, but focus only on the temporal dynamics of odorant concentration induced by turbulence, which is largely independent of odorant identity. 

This study points out that obtaining clean and reliable data for gas recognition benchmarks is still very difficult. There are manifold challenges in designing and manufacturing a gas sensor setup, as well as in planning a recording campaign robust against undesired sensor properties like drift. As shown here, the design of the experiment could hold pitfalls that may not be evident from the outset. Researchers relying on third-party datasets are therefore well advised to check the validity of a gas-sensor dataset before using it as a basis to develop algorithms for gas sensing.

A few recommendations emerge from our analysis towards best practices for designing MOx gas sensor datasets and sharing them. First and foremost, it is imperative to use a reference gas at short time intervals that will allow the identification and quantification of deviations in sensor response. Second, individual gases or mixtures should ideally be presented in pseudo-randomized order, as should any parameters that are varied (e.g.\ wind speed, hotplate voltage). If a randomized presentation order is not feasible, one should record multiple batches for the same set of parameters. Doing so, cross-validation train and test splits could be selected from batches that are time-separated (as in \cite{ASADI2017157}), which would allow for a more realistic performance evaluation. Finally, external parameters that could affect sensor behavior should be measured and reported, e.g.\ ambient temperature and humidity, and the exact time of the recording. 

Reliable data is the foundation for the development of robust and performant algorithms for gas sensing. The large number of citations of the original publication of the dataset analyzed here indicates that such data is much sought after and of high value for the community. It underlines the requirement for future efforts to record and publicly share gas sensing data for the progress of the field as a whole.

\newpage
\section{Author Contribution Statement}
\textbf{Nik Dennler:} Conceptualization, Investigation, Formal Analysis, Software, Visualization, Writing – Original Draft, Writing – Review \& Editing. \textbf{Shavika Rastogi:} Conceptualization, Investigation, Validation, Writing – Review \& Editing. \textbf{Jordi Fonollosa:} Writing – Review \& Editing. \textbf{André van Schaik:} Conceptualization, Writing – Review \& Editing, Supervision. \textbf{Michael Schmuker:} Conceptualization, Writing – Review \& Editing, Supervision. 

\section{Acknowledgements}
\label{sec:acknowledgements}
We thank A.\ J.\ Lilienthal for fruitful discussions at multiple occasions, which led to valuable insights. MS was funded by the NSF/CIHR/DFG/FRQ/UKRI-MRC Next Generation Networks for Neuroscience Program (NSF award no.\ 2014217, MRC award no. MR/T046759/1), and the EU flagship Human Brain Project SGA3 (H2020 award no.\ 945539). JF acknowledges the Spanish Ministry of Economy and Competitiveness DPI2017-89827-R, Networking Biomedical Research Centre in the subject area of Bioengineering, Biomaterials and Nanomedicine, initiatives of Instituto de Investigación Carlos III, Share4Rare project (grant agreement 780262), and ACCIÓ (Innotec ACE014/20/000018). JF also acknowledges the CERCA Programme/Generalitat de Catalunya and the Serra Húnter Program. B2SLab is certified as 2017 SGR 952. 

\section{Code Availability Statement}
Code for data cleaning and analysis will be uploaded to a publicly available repository once the paper has gone through peer-review and has been conditionally accepted.



 \bibliographystyle{elsarticle-num} 
 \bibliography{references}





\end{document}